\documentclass[aps,preprint,showpacs,preprintnumbers,amsmath,amssymb]{revtex4}
\usepackage{amsmath,mathrsfs,amsbsy,color,graphicx,bm,amsthm,amsfonts}
\usepackage{units}
\usepackage{bbm}
\usepackage{times}
\usepackage{dcolumn}
\usepackage{mathrsfs}
\usepackage{amsmath,amssymb,epsfig}
\usepackage{amsmath}
\newcommand{\udots}{\mathinner{\mskip1mu\raise1pt\vbox{\kern7pt\hbox{.}}
\mskip2mu\raise4pt\hbox{.}\mskip2mu\raise7pt\hbox{.}\mskip1mu}}
\begin{document}

\title{ Does acceleration always degrade quantum entanglement  for tetrapartite  Unruh-DeWitt detectors?  }
\author{Si-Han Li, Si-Han Shang, Shu-Min Wu\footnote{Email: smwu@lnnu.edu.cn}}
\affiliation{ Department of Physics, Liaoning Normal University, Dalian 116029, China}


\begin{abstract}
Previous studies have shown that the Unruh effect completely destroys quantum entanglement and coherence of bipartite states, as modeled by entangled Unruh-DeWitt detectors. But does the Unruh effect have a different impact on quantum entanglement of multipartite states within this framework? In this paper, we investigate the influence of the Unruh effect on $1-3$ entanglement in the context of entangled tetrapartite Unruh-DeWitt detectors. We find that quantum entanglement of tetrapartite $W$ state first decreases to a minimum value and then increases to a fixed value with the growth of the acceleration. This indicates that the Unruh effect can, under certain conditions, enhance quantum entanglement. In other words, the Unruh effect plays a dual role in the behavior of quantum entanglement-both diminishing and enhancing it. This discovery challenges and overturns the traditional view that the Unruh effect is solely detrimental to quantum entanglement and coherence in entangled Unruh-DeWitt detectors, offering a fresh and profound perspective on its impact.
\end{abstract}

\vspace*{0.5cm}
\maketitle
\section{Introduction}
Quantum entanglement is a cornerstone of quantum mechanics, characterized by the inseparability of tensor product states involving two or more particles.
Recognized as a crucial physical resource, it underpins essential tasks such as quantum teleportation \cite{M112,M113,M114,M2,M3}, quantum computing \cite{M115,M5,M6}, quantum cryptography \cite{M116,M10}, dense coding \cite{M117,M118,M119}, and quantum communication \cite{M120,M121,M12,M13}. Consequently, the study of quantum entanglement has attracted significant attention from the scientific community.  Among its forms, multipartite entanglement stands out as a vital resource in quantum information processing \cite{M123,M21,M22,M23}, offering distinct advantages over bipartite entanglement and enabling a range of physical applications. However, its complexity grows exponentially with the number of particles, making the experimental preparation and manipulation of such states significantly more challenging. Theoretical analyses of multipartite entanglement also demand advanced mathematical frameworks and substantial computational resources, further complicating research in this area. Despite these challenges, the unique properties and benefits of multipartite entanglement, particularly in relativistic contexts, underscore its importance in advancing quantum science.

Relativistic quantum information is a rapidly evolving interdisciplinary field that brings together the principles of quantum information theory, quantum field theory, and general relativity.  With significant advancements in theory \cite{R1,R2,R3,R4,R5,R6,R7,R8,R9,R10,R11,R12,R13,R14,R15,R16,R17,R18,RM1,RM2,RM3,RM4,RM5,RM6,RM7,RM8,RM9,RM10,RM11,RM12,QQRM12,WDD1,WDD2,rbn1,rbn2}, simulation \cite{SI1,SI2,SI3,SI4,SI5,SI6,SI7}, and experiment \cite{EI1,EI2,EI3,EI4}, this discipline is paving the way for deeper insights into the nature of quantum phenomena in relativistic context. Theoretically, studies have shown that both bipartite entanglement and coherence under relativistic background decrease with the increase of the acceleration (or Hawking temperature) within the free field mode model of single-mode approximation  \cite{R1,R2,R3,R4,R5,R6,R7,R8,R9,R10,R11,R12,R13,R14,R15,R16,R17,R18}.
Decoherence has also been observed in multipartite entanglement and coherence under relativistic effects  \cite{RM1,RM2,RM3,RM4,RM5,RM6,RM7,RM8,RM9,RM10,RM11,RM12}.
However, while the free field mode model is crucial for theoretical exploration, it lacks practical relevance for experimental verification.  This gap is bridged by the Unruh-Dewitt detector model, which offers a more realistic representation by employing two-level, semiclassical atoms with a fixed energy gap.
These detectors interact locally with neighboring scalar fields, thereby overcoming the impracticality of detecting global free models in the full space \cite{UD1,UD2,UD3,UD4,UD5,UD6,UD7,UD8,UD9,UD10,UD11,UD12,UD13}. Within this framework, quantum steering, entanglement, discord, and coherence between a pair of entangled Unruh-Dewitt detectors are disrupted by the Unruh effect, particularly when one detector undergoes acceleration \cite{UD14,UD15,UD16,UD17,UD18,UD19}.
Therefore, one of the motivations is to investigate whether quantum entanglement of multipartite state is more robust against the Unruh effect compared to quantum entanglement of bipartite state in the context of entangled Unruh-DeWitt detectors. Additionally, since the Unruh effect uniformly degrades quantum entanglement and coherence in both bipartite and multipartite systems under the single-mode approximation \cite{R1,R2,R3,R4,R5,R6,R7,R8,R9,R10,R11,R12,R13,R14,R15,R16,R17,R18,RM1,RM2,RM3,RM4,RM5,RM6,RM7,RM8,RM9,RM10,RM11,RM12}, another motivation is to explore whether this effect similarly influences multipartite entanglement in Unruh-DeWitt detector systems.

Based on the above motivations, in this paper,  we investigate the dynamics of quantum entanglement ($1-3$ tangle) in $W$ and $GHZ$ states within an entangled tetrapartite relativistic system, where one Unruh-DeWitt detector undergoes acceleration. The detector is treated as classical in terms of its worldline but quantum in terms of its internal degrees of freedom, which are modeled quantum mechanically.
We find that unlike quantum entanglement of bipartite state, which is completely destroyed by the Unruh effect \cite{UD14,UD15,UD16,UD17,UD18,UD19}, quantum entanglement of tetrapartite $W$ state can persist even in the limit of infinite acceleration.  We also find that quantum entanglement of tetrapartite $W$ state does not decrease monotonically with increasing acceleration. However, the relationship between acceleration and quantum entanglement is monotonic for bipartite Unruh-DeWitt detectors model. This means that the Unruh effect plays a dual role, both reducing and increasing quantum entanglement of $W$ state  based on tetrapartite Unruh-DeWitt detectors model. These findings provide new insights into the role of the Unruh effect in quantum resources and their practical applications.

The structure of the paper is as follows. In Section II, we introduce the quantum information framework for entangled Unruh-DeWitt detectors and analyze the evolution of  tetrapartite $W$ state when one detector undergoes acceleration. In Section III, we investigate the behavior of quantum entanglement in relativistic tetrapartite quantum systems. Finally, the last section provides a summary of our findings.

\section{Evolution of tetrapartite quantum system of $W$ state with an
accelerated atom }
In this paper, we explore the dynamics of tetrapartite entangled Unruh-DeWitt detectors, focusing primarily on the initial four-qubit states: $W$ and $GHZ$ states. While this section concentrates on the $W$ state, a detailed analysis of the $GHZ$ state is provided in the Appendix A. The four-qubit $W$ state is defined as
\begin{eqnarray}\label{qq7}
|W_{4}\rangle=\frac{1}{2}(|0_A0_B0_C1_D\rangle+|0_A0_B1_C0_D\rangle+|0_A1_B0_C0_D\rangle+|1_A0_B0_C0_D\rangle).
\end{eqnarray}
For simplicity, subscripts will be omitted in the following discussion. The model consists of four observers-Alice, Bob, Charlie, and David-each equipped with a
two-level Unruh-DeWitt detector. David's detector undergoes uniform acceleration along the $x$-axis for a finite proper time interval, while the detectors of Alice, Bob, and Charlie remain stationary.  For simplicity, we assume that Alice, Bob, and Charlie's detectors are always switched off, whereas David's detector remains active due to its constant acceleration. The worldline of David's detector is parametrized by
\begin{eqnarray}\label{w66}
t(\tau)=a^{-1}\sinh a\tau, \quad x(\tau)=a^{-1}\cosh a\tau,\quad y(\tau)=z(\tau)=0,
\end{eqnarray}
where $a$ denotes David's proper acceleration, and $\tau$ represents the proper time of the detector  \cite{UD14,UD15}. Throughout this paper, we set $c=\hbar=k_{B}=1$
for simplicity. The initial state of the detector-field system is assumed to be
\begin{eqnarray}\label{A7}
|W_{4-\infty}\rangle=|W_{4}\rangle\otimes|0_{M}\rangle,
\end{eqnarray}
where $|W_{4}\rangle$ is the initial tetrapartite entangled state as defined in  Eq.(\ref{qq7}), and $|0_{M}\rangle$ represents the Minkowski vacuum of the external scalar field. The interaction between the qubit and the massless scalar field
$\phi(x)$ is modeled by the interaction Hamiltonian $H_{int}^{D\phi}(\tau)$:
\begin{eqnarray}\label{A8}
H_{int}^{D\phi}(\tau)=\epsilon(\tau)\int_{\sum_{\tau}}d^{3} \boldsymbol{x}\sqrt{-g}\phi(x)[\psi(\boldsymbol{x})D+\bar{\psi}(\boldsymbol{x})D^{\rm \dagger}],
\end{eqnarray}
where $D$ and $D^{\rm \dagger}$ are the annihilation and creation operators associated with David's detector, respectively. The coupling constant
$\epsilon(\tau)$ ensures that the detectors are active only for a finite proper time interval  $\vartriangle$, and remain inactive beyond this period. Here,  $g\equiv \det(g_{ab})$, where $g_{ab}$ is the Minkowski spacetime metric, and $\sum_{\tau=\rm{const}}$ indicates that the integration is performed over the global spacelike Cauchy surface \cite{UD19,UD20}. The function $\psi(\boldsymbol{x})=(\kappa\sqrt{2\pi})^{-3}\exp(-\boldsymbol{x}^{2}/(2\kappa^{2}))$  is a Gaussian coupling function with variance
$\kappa=\rm{const}$, signifying that the detector interacts exclusively with its neighboring field. Finally, the total Hamiltonian of the entire tetrapartite system can be expressed as
\begin{eqnarray}\label{A9}
H_{4\phi}=H_{A}+H_{B}+H_{C}+H_{D}+H_{KG}+H_{int}^{D\phi},
\end{eqnarray}
where $H_{T} = \Omega T^{\rm \dagger}T$  $(T=A,B,C,D)$  is the free Hamiltonian of each particle detector, with the energy gap $\Omega$ and $H_{KG}$ is the Hamiltonian of the massless scalar field. The creation and annihilation operators $T^{\rm \dagger}$ and $T$ satisfy the usual commutation relations: $T^{\rm \dagger}|1\rangle = T|0\rangle=0$, $T|1\rangle = |0\rangle$, and $T^{\rm \dagger}|0\rangle = |1\rangle$, where $|1\rangle$ and $|0\rangle$ are the excited and ground states of the detector, respectively.

In the interaction picture,  the final state $|W_{4\infty}\rangle$ describing the total system at the first-order perturbation is given by
\begin{eqnarray}\label{qq1}
|W_{4\infty}\rangle=(I+a^{\rm \dagger}_{RI}(\lambda)D-a_{RI}(\bar{\lambda})D^{\rm \dagger})|W_{4-\infty}\rangle,
\end{eqnarray}
where $|W_{4-\infty}\rangle$ represents the initial state, $\lambda=-KEf$,
the operators $a^{\rm \dagger}_{RI}$ and $a_{RI}$ are creation and annihilation operators of $\lambda$ modes in the Rindler region $I$, respectively.
The function $f$ is defined as $f\equiv\epsilon(t)e^{-i\Omega t}\psi(\boldsymbol{x})$.
The operator  $K$ establishes the correspondence between the positive frequency part of the solutions of the Klein-Gordon equation $\nabla_{a}\nabla^{a}\phi(x)=0$ and the timelike isometry. The $Ef$ can be written as
\begin{eqnarray}\label{qq2}
Ef=\int d^{4}x^{\prime}\sqrt{-g(x^{\prime})}[G^{\rm{adv}}(x,x^{\prime})-G^{\rm{ret}}(x,x^{\prime})]f(x^{\prime}),
\end{eqnarray}
where $E$ denotes the difference between the advanced Green's function $G^{\rm{adv}}$ and the retarded Green's function $G^{\rm{ret}}$.

Substituting the initial state $|W_{4-\infty}\rangle$  from Eq.(\ref{A7}) into Eq.(\ref{qq1}), the final state of the total system, expressed in terms of the Rindler operators $a^{\rm \dagger}_{RI}$ and $a_{RI}$, can be given by
\begin{eqnarray}\label{qqq2}
\begin{split}
|W_{4\infty}\rangle=&|W_{4-\infty}\rangle+\frac{1}{2}[|0000\rangle\otimes(a^{\rm \dagger}_{RI}(\lambda)|0_{M}\rangle)\\
&+(|0011\rangle+|0101\rangle+|1001\rangle)\otimes(a_{RI}(\bar{\lambda})|0_{M}\rangle)],
\end{split}
\end{eqnarray}
where $a^{\rm \dagger}_{RI}(\lambda)$ and $a_{RI}(\lambda)$ are defined in Rindler region $I$, and $|0_{M}\rangle$ denotes the vacuum state in  Minkowski spacetime.
The Bogoliubov transformations between the Rindler operators and the operators annihilating the Minkowski vacuum $|0_{M}\rangle$ are given by
\begin{eqnarray}\label{qq3}
a_{RI}(\bar{\lambda})=\frac{a_{M}(\overline{F_{1\Omega}})+e^{-\pi\Omega/a}a^{\rm \dagger}_{M}(F_{2\Omega})}{(1-e^{-2\pi\Omega/a})^{1/2}},
\end{eqnarray}
\begin{eqnarray}\label{qq5}
a^{\rm \dagger}_{RI}(\lambda)=\frac{a^{\rm \dagger}_{M}(F_{1\Omega})+e^{-\pi\Omega/a}a_{M}(\overline{F_{2\Omega}})}{(1-e^{-2\pi\Omega/a})^{1/2}},
\end{eqnarray}
where $F_{1\Omega}=\frac{\lambda+e^{-\pi\Omega/a}\lambda\circ w}{(1-e^{-2\pi\Omega/a})^{1/2}}$ and $F_{2\Omega}=\frac{\overline{\lambda\circ w}+e^{-\pi\Omega/a}\bar{\lambda}}{(1-e^{-2\pi\Omega/a})^{1/2}}$.
In $F_{1\Omega}$ and $F_{2\Omega}$, $w(t,x,y,z)=(-t,-x,y,z)$ represents a wedge reflection isometry, which maps the function $\lambda$ from Rindler region $I$ to $\lambda\circ w$ in Rindler region $II$, where the symbol $\circ$ denotes the composition of mappings, i.e., the composite mapping of two functions.

Utilizing the Bogliubov transformations presented in Eqs.(\ref{qq3}) and (\ref{qq5}), and adhering to the relations: $a_{M}|0_{M}\rangle=0$ and $a_{M}^{\rm \dagger}|0_{M}\rangle=|1_{M}\rangle$, Eq.(\ref{qqq2}) can be rewritten as
\begin{eqnarray}\label{qqq5}
|W_{4\infty}\rangle=|W_{4-\infty}\rangle+\frac{1}{2}\nu\left[\frac{|0000\rangle\otimes|1_{\tilde{F}_{1\Omega}}\rangle}{(1-e^{-2\pi\Omega/a})^{1/2}}
+e^{-\pi\Omega/a}\frac{(|0011\rangle+|0101\rangle+|1001\rangle)\otimes|1_{\tilde{F}_{2\Omega}}\rangle}{(1-e^{-2\pi\Omega/a})^{1/2}}\right],
\end{eqnarray}
where $\tilde{F}_{\rm{i}\Omega}=F_{\rm{i}\Omega}/\nu$.  To obtain the dynamical state of the detectors after interaction with the field, we trace out the external field degrees of freedom, yielding the reduced density matrix for the four-qubit state
\begin{eqnarray}
\rho^{{ABCD}}_{\infty(W)}=\|W_{4\infty}\|^{-2}{\rm{tr}}_{\phi}|W_{4\infty}\rangle\langle W_{4\infty}|,
\end{eqnarray}
where $\|W_{4\infty}\|^{2}$ ensures the normalization  and is given by
$$\|W_{4\infty}\|^{2}=1+\frac{\nu^{2}(1+3e^{-2\pi\Omega/a})}{4(1-e^{-2\pi\Omega/a})}.$$
Thus, the final state of the detectors is
\begin{eqnarray}\label{pp6}
\rho^{{ABCD}}_{\infty(W)}=\rho_{diag}^{W_{4}}+\rho_{nondiag}^{W_{4}},
\end{eqnarray}
with
\begin{eqnarray}\label{ppp6}
\begin{split}
\rho_{diag}^{W_{4}}=&K_{0}|0000\rangle\langle0000|+K_{1}(|0001\rangle\langle0001|+|0010\rangle\langle0010|+|0100\rangle\langle0100|\\
&+|1000\rangle\langle1000|)+K_{2}(|0011\rangle\langle0011|+|0101\rangle\langle0101|+|1001\rangle\langle1001|),
\end{split}
\end{eqnarray}
and
\begin{eqnarray}\label{pppp6}
\begin{split}
\rho_{nondiag}^{W_{4}}=&K_{1}(|0001\rangle\langle0010|+|0001\rangle\langle0100|+|0001\rangle\langle1000|+|0010\rangle\langle0100|\\
&+|0010\rangle\langle1000|+|0100\rangle\langle1000|)+K_{2}(|0011\rangle\langle0101|+|0011\rangle\langle1001|\\
&+|0101\rangle\langle1001|)+(H.c.)_{nondiag.},
\end{split}
\end{eqnarray}
where $K_{0}=\frac{\nu^{2}}{4(1-q)+\nu^{2}(1+3q)}$, $K_{1}=\frac{1-q}{4(1-q)+\nu^{2}(1+3q)}$, and $K_{2}=\frac{\nu^{2}q}{4(1-q)+\nu^{2}(1+3q)}$.
Here, the parametrized acceleration $q$ is defined as $q\equiv e^{-2\pi\Omega/a}$, and the effective coupling parameter $\nu^{2}\equiv\|\lambda\|^{2}=\frac{\epsilon^{2}\Omega\vartriangle}{2\pi}e^{-\Omega^{2}\kappa^{2}}$. Moreover, these expressions are valid under the conditions $\epsilon\ll\Omega^{-1}\ll\vartriangle$ and $\epsilon$ is a slowly varying function of time compared to the frequency $\Omega$.
To ensure the validity of the perturbative approach, the coupling parameter is restricted to $\nu^{2}\ll1$. Additionally, the parameter $q$ behaves as a monotonic function of acceleration, with extreme limits  $q\rightarrow0$ corresponding to the zero acceleration and $q\rightarrow1$ corresponding to the infinite acceleration.

\section{Behaviors of quantum entanglement of tetrapartite $W$ and $GHZ$ states under the influence of Unruh thermal noise}
Negativity is a widely used measure for quantifying entanglement in various quantum systems. It determines whether a system remains entangled by examining the presence of negative eigenvalues in the partial transpose of its density matrix. Specifically, a state is considered entangled if at least one negative eigenvalue exists in the partial transpose. For a tetrapartite state, the negativity is defined as
\begin{eqnarray}\label{TH1}
N_{\alpha(\beta\gamma\eta)}=\|\rho^{{T_{\alpha}}}_{\alpha\beta\gamma\eta}\|-1,
\end{eqnarray}
which describes  $1-3$ tangle. Here, $T_{\alpha}$ is the partial transpose of $\rho_{\alpha\beta\gamma\eta}$ with respect to the observer $\alpha$ \cite{NE1}. The term $\|A\|-1 $ refers to twice the sum of the absolute values of the negative eigenvalues of the operator $A$.  Consequently, the negativity can also be expressed as
\begin{eqnarray}\label{TH2}
N_{\alpha(\beta\gamma\eta)}=2\sum^{n}_{i=1}|\lambda^{(-)}_{\alpha(\beta\gamma\eta)}|^{i},
\end{eqnarray}
where $|\lambda^{(-)}_{\alpha(\beta\gamma\eta)}|^{i}$  denotes the negative eigenvalues of the partial transpose matrix. This formulation provides a clear and efficient method for calculating the entanglement measure.

We investigate the negativity ($1-3$ tangle) for tetrapartite $W$ state in density operator $\rho^{{ABCD}}_{\infty(W)}$. To achieve this, we first compute the partial transpose of $\rho^{{ABCD}}_{\infty(W)}$ with respect to $A$ mode, leading to the expression
\begin{eqnarray}\label{TH3}
\rho^{{T_{A}}}_{\infty(W)}=\rho_{diag}^{W_{4}}+\rho_{nondiag}^{W_{4}(T_{A})},
\end{eqnarray}
with
\begin{eqnarray}\label{qq6}
\begin{split}
\rho_{nondiag}^{W_{4}(T_{A})}=&K_{1}(|0000\rangle\langle1001|+|0000\rangle\langle1010|+|0000\rangle\langle1100|+|0001\rangle\langle0010|\\
&+|0001\rangle\langle0100|+|0010\rangle\langle0100|)+K_{2}(|0001\rangle\langle1011|+|0001\rangle\langle1101|\\
&+|0011\rangle\langle0101|)+(H.c.)_{nondiag.}.
\end{split}
\end{eqnarray}
Due to the complexity of the expression for $N_{A(BCD)}(W_4)$, we refrain from presenting it explicitly here. Similarly, by taking the transpose with respect to the mode $D$, we obtain
\begin{eqnarray}\label{TH4}
\rho^{{T_{D}}}_{\infty(W)}=\rho_{diag}^{W_{4}}+\rho_{nondiag}^{W_{4}(T_{D})},
\end{eqnarray}
with
\begin{eqnarray}\label{qqq6}
\begin{split}
\rho_{nondiag}^{W_{4}(T_{D})}=&K_{1}(|0000\rangle\langle0011|+|0000\rangle\langle0101|+|0000\rangle\langle1001|+|0010\rangle\langle0100|\\
&+|0010\rangle\langle1000|+|0100\rangle\langle1000|)+K_{2}(|0011\rangle\langle0101|+|0011\rangle\langle1001|\\
&+|0101\rangle\langle1001|)+(H.c.)_{nondiag.}.
\end{split}
\end{eqnarray}
By employing the Eq.(\ref{TH2}), the negativity of the tetrapartite $W$ state can be expressed as
\begin{eqnarray}\label{TH5}
N_{D(ABC)}(W_4)=\max\left\{0,\frac{-\nu^{2}-3q\nu^{2}+\sqrt{12-24q+12q^{2}+\nu^{4}-6q\nu^{4}+9q^{2}\nu^{4}}}{4-4q+\nu^{2}+3q\nu^{2}}\right\}.
\end{eqnarray}

\begin{figure}
\begin{minipage}[t]{0.5\linewidth}
\centering
\includegraphics[width=3.0in,height=5.2cm]{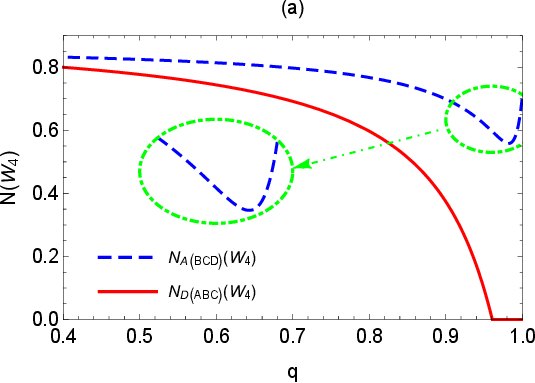}
\label{fig1a}
\end{minipage}%
\begin{minipage}[t]{0.5\linewidth}
\centering
\includegraphics[width=3.0in,height=5.2cm]{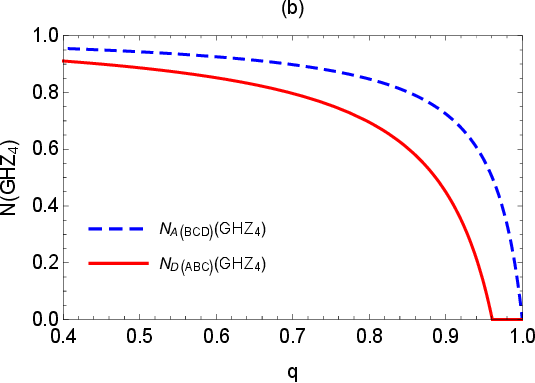}
\label{fig1b}
\end{minipage}%
\caption{The $N_{A(BCD)}$ and $N_{D(ABC)}$ of $W$ and $GHZ$ states as a function of the acceleration parameter $q$. The effective coupling parameter $\nu$ is fixed as $\nu^{2}=0.04$. }
\label{Fig1}
\end{figure}

In Fig.\ref{Fig1}, we present the behavior of the negativity ($1-3$ tangle) of tetrapartite $W$ and $GHZ$ states as a function of the acceleration parameter  $q$, while keeping the effective coupling parameter fixed. For a detailed calculation of the negativity in the $GHZ$ state, please see Appendix A. As shown in Fig.\ref{Fig1}(a), quantum entanglement $N_{A(BCD)}(W_4)$ of $W$ state exhibits a non-monotonic behavior: it initially decreases to a minimum value before increasing again to a fixed value with the rise of the acceleration parameter $q$. This suggests  that there is a
non-monotonic relationship between quantum entanglement $N_{A(BCD)}(W_4)$ and acceleration.  In contrast, quantum entanglement and coherence of bipartite states  decrease monotonically to zero with increasing acceleration parameter $q$, which shows that the Unruh effect completely destroys them \cite{UD14,UD15,UD16,UD17}. From this analysis, we observe that the Unruh effect acts solely as a decoherence mechanism for bipartite entanglement and coherence. However, its impact on the quantum entanglement
$N_{A(BCD)}(W_4)$ of $W$ state  is more nuanced, acting as a double-edged sword that can both inhibit and promote entanglement. This insight provides a deeper and more comprehensive understanding of the Unruh effect's influence on quantum resources. Additionally, Fig.\ref{Fig1} reveals that, unlike quantum entanglement $N_{A(BCD)}(W_4)$ subject to Unruh effect,  $N_{D(ABC)}(GHZ_4)$, $N_{A(BCD)}(GHZ_4)$, and $N_{D(ABC)}(W_4)$ are completely destroyed by the Unruh effect.

\begin{figure}
\begin{minipage}[t]{0.5\linewidth}
\centering
\includegraphics[width=3.0in,height=5.2cm]{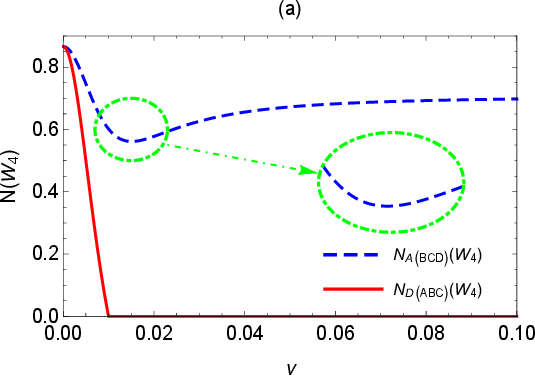}
\label{fig2a}
\end{minipage}%
\begin{minipage}[t]{0.5\linewidth}
\centering
\includegraphics[width=3.0in,height=5.2cm]{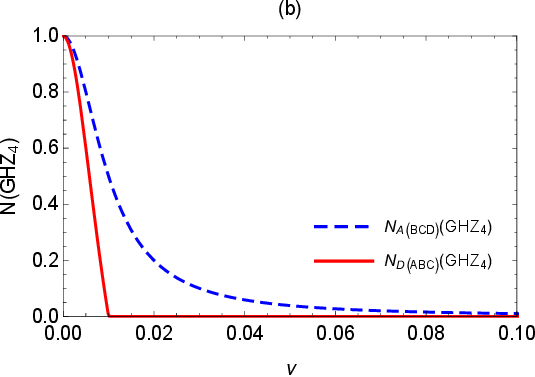}
\label{fig2b}
\end{minipage}%
\caption{The $N_{A(BCD)}$ and $N_{D(ABC)}$ of $W$ and $GHZ$ states as a function of the effective coupling parameter $\nu$ with a fixed acceleration parameter $q=0.9999$. }
\label{Fig2}
\end{figure}

Fig.\ref{Fig2} illustrates the behavior of negativity as a function of the effective coupling parameter $\nu$ for two distinct types of tetrapartite entangled states: $W$ and $GHZ$ states, under extreme acceleration ($q=0.9999$). It is evident that quantum entanglement, specifically $N_{D(ABC)}(W_{4})$,   $N_{A(BCD)}(GHZ_4)$, and $N_{D(ABC)}(GHZ_4)$, initially undergoes a sharp decline before experiencing ``sudden death" with the increasing value of $\nu$. This behavior indicates that the interaction between the detector and the field leads to a loss of quantum entanglement, effectively transferring the entanglement from the detectors to the detector-field system. In other words, the entanglement that was initially present among the detectors is now redistributed between the detectors and the field.
Interestingly, quantum entanglement $N_{A(BCD)}(W_4)$ of the tetrapartite $W$ state follows a different pattern: it initially decreases, reaches a minimum, and then gradually recovers, eventually stabilizing at a fixed value as the coupling parameter $\nu$ increases. This recovery suggests that the entanglement between the detector and the field can be transferred back into entanglement between the detectors, revealing the dynamic and reversible nature of entanglement in multipartite systems. Unlike the bipartite Unruh-Dewitt detector model, where information typically flows from the detectors to the field \cite{UD14,UD15,UD16,UD17}, in the multipartite model, the flow of information is bidirectional, with the potential for entanglement to return from the field to the detectors.
This provides valuable insight into how entanglement can evolve and be redistributed in the presence of strong interactions with the field, offering new possibilities for controlling and manipulating entanglement in quantum systems.

\begin{figure}
\begin{minipage}[t]{0.5\linewidth}
\centering
\includegraphics[width=3.0in,height=6.24cm]{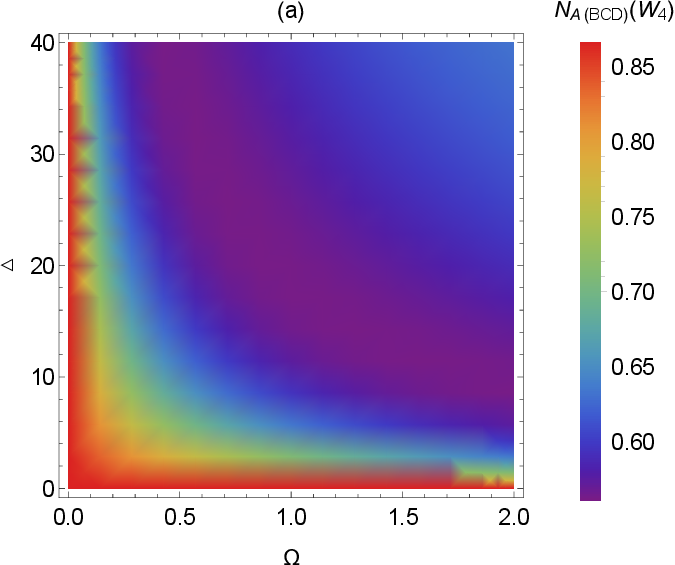}
\label{fig3a}
\end{minipage}%
\begin{minipage}[t]{0.5\linewidth}
\centering
\includegraphics[width=3.0in,height=6.24cm]{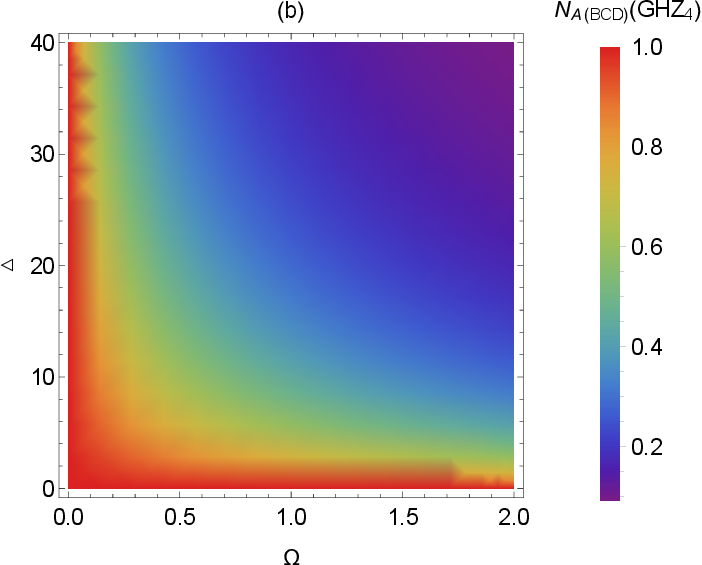}
\label{fig3b}
\end{minipage}%
\caption{The  $N_{A(BCD)}$ of the tetrapartite systems for $W$ and $GHZ$ states as functions of the interaction time duration $\vartriangle$ and the energy gap $\Omega$, with $\epsilon^{2}=8\pi^{2}\cdot10^{-6}$, $\kappa=0.02$, and $q=0.9999$. }
\label{Fig3}
\end{figure}

We further explore how the interaction between the accelerated detector and the external scalar field affects the negativity in the tetrapartite $W$ and $GHZ$ states by plotting the negativity $N_{A(BCD)}$  as a function of interaction time duration $\vartriangle$ and the energy gap $\Omega$ in Fig.\ref{Fig3}.
The results reveal a striking difference between the two states. While $N_{A(BCD)}(W_4)$ exhibits a non-monotonic trend in response to both $\vartriangle$ and $\Omega$, $N_{A(BCD)}(GHZ_4)$ shows a monotonic decrease as these parameters increase. This indicates that the $W$ state exhibits a certain degree of resilience against the Unruh effect, with the possibility of partially recovering its entanglement by redistributing quantum correlations between the detector system and the scalar field. In contrast, the $GHZ$ state undergoes a rapid and irreversible loss of entanglement under the same conditions. These findings underscore the significant differences in the behavior of $W$ and $GHZ$ states in a relativistic context. They highlight that the structural properties of these states play a pivotal role in determining their robustness against relativistic effects, such as the Unruh effect. As a result, quantum entanglement of  $W$ state emerges as a more favorable candidate for processing relativistic quantum information tasks. By capitalizing on the unique properties of $W$ state entanglement, it is possible to design advanced detectors, using artificial two-level atoms with carefully chosen energy gaps, to preserve quantum entanglement even in the presence of Unruh-induced thermal noise. This research opens new avenues for robust quantum information processing in relativistic environments.

\section{Conclusions}
In this study, we have explored the significant influence of the Unruh effect on $1-3$ entanglement of $W$ and $GHZ$ states within tetrapartite systems, specifically when one of the detectors undergoes uniform acceleration. Our results reveal striking contrasts in the impact of the Unruh effect on the entanglement of different types of states. Notably, we find that quantum entanglement of tetrapartite $W$ state remains remarkably resilient to the Unruh effect, while the entanglement of tetrapartite $GHZ$ state is completely destroyed. Similarly, the Unruh effect leads to the degradation of both entanglement and coherence in bipartite states, as established in previous studies \cite{UD14, UD15, UD16, UD17}.
Intriguingly, we uncover a dual nature of the Unruh effect on entanglement of $W$ state  within the Unruh-Dewitt detector framework. It can both degrade and enhance entanglement, depending on the specifics of the interaction, thus acting as a double-edged sword. This stands in stark contrast to its unidirectional role in bipartite states, where it solely induces decoherence. This novel observation challenges the conventional wisdom that the Unruh effect merely destroys quantum resources such as entanglement and coherence for Unruh-DeWitt detectors mode. Instead, it highlights the potential for the Unruh effect to play a more complex, nuanced role in quantum information processes.  Additionally, we observe an intriguing non-monotonic behavior in the entanglement of $W$ state over the interaction time, which differs from the monotonic behavior typically seen in bipartite entanglement. This non-monotonicity means that, under certain conditions, quantum entanglement can be extracted from quantum field into detectors. These findings are crucial for advancing the application of $W$ state entanglement in relativistic quantum information tasks, and may pave the way for innovative quantum technologies that leverage the interplay between the Unruh effect and quantum entanglement.

\begin{acknowledgments}
This work is supported by the National Natural
Science Foundation of China (Grant Nos. 12205133) and  the Special Fund for Basic Scientific Research of Provincial Universities in Liaoning under grant NO. LS2024Q002.	
\end{acknowledgments}


\appendix
\onecolumngrid
\section{ Dynamic quantum entanglement of tetrapartite $GHZ$ state for the Unruh-DeWitt detector model }
In this section, we present the derivation of $GHZ$ state in the evolution of the tetrapartite
Unruh-DeWitt detector system. We assume that among the tetrapartite  detectors, only David's detector moves with uniform acceleration, while the other three remain stationary.
Initially, Alice, Bob, Charlie, and David share a $GHZ$ state in Minkowski spacetime, given by
\begin{eqnarray}\label{a7}
|{GHZ}_{4-\infty}\rangle=|{GHZ}_{4}\rangle\otimes|0_{M}\rangle,
\end{eqnarray}
with
\begin{eqnarray}\label{W6}
|{GHZ}_{4}\rangle=\frac{1}{\sqrt{2}}(|0000\rangle+|1111\rangle).
\end{eqnarray}
In the interaction picture, considering the first-order perturbation, the final state of the detector-field system is determined by
\begin{eqnarray}\label{QQ1}
|{GHZ}_{4\infty}\rangle=(I+a^{\rm \dagger}_{RI}(\lambda)D-a_{RI}(\bar{\lambda})D^{\rm \dagger})|{GHZ}_{4-\infty}\rangle.
\end{eqnarray}

Substituting the initial state $|{GHZ}_{4-\infty}\rangle$ from Eq.(\ref{a7}) into Eq.(\ref{QQ1}), the final state of the total system, expressed in terms of the Rindler operators $a^{\rm \dagger}_{RI}$ and $a_{RI}$, can be written as
\begin{eqnarray}\label{QQ2}
|{GHZ}_{4\infty}\rangle=|{GHZ}_{4-\infty}\rangle+\frac{1}{\sqrt{2}}[|1110\rangle\otimes(a^{\rm \dagger}_{RI}(\lambda)|0_{M}\rangle)+|0001\rangle\otimes(a_{RI}(\bar{\lambda})|0_{M}\rangle)],
\end{eqnarray}
where the creation and annihilation operators $a^{\dagger}_{RI}(\lambda)$ and $a_{RI}(\overline{\lambda}) $ are defined in the Rindler region $I$, and $|0_{M}\rangle$ is the Minkowski vacuum. Using the Bogoliubov transformations provided in Eqs.(\ref{qq3}) and (\ref{qq5}), along with the relations: $a_{M}|0_{M}\rangle=0$ and $a_{M}^{\rm \dagger}|0_{M}\rangle=|1_{M}\rangle$, Eq.(\ref{QQ2}) can be reformulated as
\begin{eqnarray}\label{QQQ5}
|{GHZ}_{4\infty}\rangle=|{GHZ}_{4-\infty}\rangle+\frac{1}{\sqrt{2}}\nu\left[\frac{|1110\rangle\otimes|1_{\tilde{F}_{1\Omega}}\rangle}{(1-e^{-2\pi\Omega/a})^{1/2}}+e^{-\pi\Omega/a}\frac{|0001\rangle\otimes|1_{\tilde{F}_{2\Omega}}\rangle}{(1-e^{-2\pi\Omega/a})^{1/2}}\right],
\end{eqnarray}
where $\widetilde{F}=F_{i\Omega}/\nu$.

To obtain the evolution of the detectors' state after their interaction with the field, we trace out the scalar field degrees of freedom, yielding the reduced density matrix for the four-qubit state
\begin{eqnarray}\label{Q6}
\rho^{{ABCD}}_{\infty(G)}=\|{GHZ}_{4\infty}\|^{-2}{\rm{tr}}_{\phi}|{GHZ}_{4\infty}\rangle\langle {GHZ}_{4\infty}|,
\end{eqnarray}
where $\|{GHZ}_{4\infty}\|^{2}=1+\frac{\nu^{2}(1+e^{-2\pi\Omega/a})}{2(1-e^{-2\pi\Omega/a})}$, ensures the final density matrix is normalized, i.e., $\rm{tr}\rho^{{ABCD}}_{\infty(G)}=1$.
Thus, the final state of the detectors is shown to be
\begin{eqnarray}\label{QQ6}
\begin{split}
\rho^{{ABCD}}_{\infty(G)}=&L_{0}(|0000\rangle\langle0000|+|0000\rangle\langle1111|+|1111\rangle\langle0000|+|1111\rangle\langle1111|)\\
&+L_{1}|0001\rangle\langle0001|+L_{2}|1110\rangle\langle1110|,
\end{split}
\end{eqnarray}
where the parameters $L_{0}$, $L_{1}$, and $L_{2}$ are defined as follows: $L_{0}=\frac{1-q}{2(1-q)+\nu^{2}(1+q)}$, $L_{1}=\frac{\nu^{2}q}{2(1-q)+\nu^{2}(1+q)}$, and $L_{2}=\frac{\nu^{2}}{2(1-q)+\nu^{2}(1+q)}$.

Taking the partial transpose of the density matrix $\rho^{ABCD}_{\infty(G)}$ with respect to mode $A$, we obtain
\begin{eqnarray}\label{QQQ6}
\begin{split}
\rho^{{T_{A}}}_{\infty(G)}=&L_{0}(|0000\rangle\langle0000|+|0111\rangle\langle1000|+|1000\rangle\langle0111|+|1111\rangle\langle1111|)\\
&+L_{1}|0001\rangle\langle0001|+L_{2}|1110\rangle\langle1110|.
\end{split}
\end{eqnarray}
By employing the Eq.(\ref{TH2}), the negativity of the tetrapartite $GHZ$ state, considering the partial transpose with respect to mode $A$, is given by
\begin{eqnarray}\label{QQ9}
N_{A(BCD)}(GHZ_4)=\max\left\{0,\frac{2(1-q)}{2-2q+\nu^{2}+q\nu^{2}}\right\}.
\end{eqnarray}
Similarly, the partial transpose with respect to mode $D$ can be expressed as
\begin{eqnarray}\label{QQ8}
\begin{split}
\rho^{{T_{D}}}_{\infty(G)}=&L_{0}(|0000\rangle\langle0000|+|0001\rangle\langle1110|+|1110\rangle\langle0001|+|1111\rangle\langle1111|)\\
&+L_{1}|0001\rangle\langle0001|+L_{2}|1110\rangle\langle1110|.
\end{split}
\end{eqnarray}
For mode $D$, the negativity takes the form
\begin{eqnarray}\label{QQQ9}
N_{D(ABC)}(GHZ_4)=\max\left\{0,\frac{-\nu^{2}-q\nu^{2}+(1-q)\sqrt{4+\nu^{4}}}{2-2q+\nu^{2}+q\nu^{2}}\right\}.
\end{eqnarray}

\section{Dynamic quantum entanglement of tetrapartite $ W $ and $GHZ$ states for the Unruh-DeWitt detector model under the second-order perturbation }
In this appendix, we investigate the dynamic quantum entanglement of the tetrapartite $W$ and $GHZ$ states for the Unruh-DeWitt detector model under the second-order perturbation.
The time-evolution operator $U$ in the interaction picture is defined using the Dyson series with the time-ordering operator $\mathcal{T}$ \cite{NE2, NE3}, and it takes the form
\begin{eqnarray}\label{S1}
\begin{split}
U &:= \mathcal{T} \exp \left[ -i \int d\tau \, H_D(\tau) \right] \\
&= 1 + (-i) \int d\tau \, H_D(\tau) + \frac{(-i)^2}{2} \int d\tau \, d\tau' \, \mathcal{T} H_D(\tau) H_D(\tau') + \mathcal{O}\left( \nu^3 \right),
\end{split}
\end{eqnarray}
where the second-order terms account for sequential interactions governed by the causal structure $\tau > \tau'$. Applying this to the initial $W_4$ state $|W_{4-\infty}\rangle$ from Eq.(\ref{A7}), the final state $|W^{(2)}_{4\infty}\rangle$ describing the total system at the second-order perturbation, expressed in terms of the Rindler operators $a^{\rm \dagger}_{RI}$ and $a_{RI}$, is given by
\begin{eqnarray}\label{S2}
\begin{split}
|W^{(2)}_{4\infty}\rangle=&[I+a^{\rm \dagger}_{RI}(\lambda)D-a_{RI}(\bar{\lambda})D^{\rm \dagger}-\frac{1}{2}(a^{\rm \dagger}_{RI}(\lambda)a_{RI}(\bar{\lambda})DD^{\rm \dagger}\\
&+a_{RI}(\bar{\lambda})a^{\rm \dagger}_{RI}(\lambda)D^{\rm \dagger}D)]|W_{4-\infty}\rangle + \mathcal{O}\left( \nu^3 \right).
\end{split}
\end{eqnarray}
Substituting the initial state $|W_{4-\infty}\rangle$ into Eq.(\ref{S2}), we obtain
\begin{eqnarray}\label{S3}
\begin{split}
|W^{(2)}_{4\infty}\rangle=&|W_{4-\infty}\rangle+\frac{1}{2}[|0000\rangle\otimes(a^{\rm\dagger}_{RI}(\lambda)|0_{M}\rangle)+(|0011\rangle+|0101\rangle+|1001\rangle)\otimes\\
&(a_{RI}(\bar{\lambda})|0_{M}\rangle)+
\frac{1}{2}(|0010\rangle+|0100\rangle+|1000\rangle)\otimes(a^{\rm\dagger}_{RI}(\lambda)a_{RI}(\bar{\lambda})|0_{M}\rangle)\\
&+\frac{1}{2}|0001\rangle\otimes(a_{RI}(\bar{\lambda})a^{\rm\dagger}_{RI}(\lambda)|0_{M}\rangle)] + \mathcal{O}\left( \nu^3 \right).
\end{split}
\end{eqnarray}
Utilizing the Bogoliubov transformations from Eqs.(\ref{qq3}) and (\ref{qq5}), and applying the relations: $a_{M}|0_{M}\rangle=0$ and $a_{M}^{\rm \dagger}|0_{M}\rangle=|1_{M}\rangle$, Eq.(\ref{S3}) can be rewritten as
\begin{eqnarray}\label{S4}
\begin{split}
|W^{(2)}_{4\infty}\rangle=&|W_{4-\infty}\rangle+\frac{1}{2}\nu\bigg[\frac{|0000\rangle\otimes|1_{\tilde{F}_{1\Omega}}\rangle}{(1-e^{-2\pi\Omega/a})^{1/2}}
+e^{-\pi\Omega/a}\frac{(|0011\rangle+|0101\rangle+|1001\rangle)\otimes|1_{\tilde{F}_{2\Omega}}\rangle}{(1-e^{-2\pi\Omega/a})^{1/2}}\\
&+\nu\frac{(|0010\rangle+|0100\rangle+|1000\rangle)\otimes(e^{-\pi\Omega/a}|1_{\tilde{F}_{1\Omega}}1_{\tilde{F}_{2\Omega}}\rangle-e^{-2\pi\Omega/a}|0_{M}\rangle)}{2(1-e^{-2\pi\Omega/a})}\\
&+\nu\frac{|0001\rangle\otimes(e^{-\pi\Omega/a}|1_{\tilde{F}_{1\Omega}}1_{\tilde{F}_{2\Omega}}\rangle-|0_{M}\rangle)}{2(1-e^{-2\pi\Omega/a})}\bigg] + \mathcal{O}\left( \nu^3 \right).
\end{split}
\end{eqnarray}
To simplify, we substitute the initial state $|W_{4-\infty}\rangle$ into Eq.(\ref{S4}), obtaining the total system's state
\begin{eqnarray}\label{S44}
\begin{split}
|W^{(2)}_{4\infty}\rangle=&\bigg[\left(\frac{1}{2}-\frac{\nu^2}{4(1-e^{-2\pi\Omega/a})}\right)|0001\rangle+\left(\frac{1}{2}-\frac{e^{-2\pi\Omega/a}\nu^2}{4(1-e^{-2\pi\Omega/a})}\right)\\
&(|0010\rangle+|0100\rangle+|1000\rangle)\bigg]\otimes|0_M\rangle+\nu\frac{|0000\rangle\otimes|1_{\tilde{F}_{1\Omega}}\rangle}{2(1-e^{-2\pi\Omega/a})^{1/2}}\\
&+e^{-\pi\Omega/a}\nu\frac{(|0011\rangle+|0101\rangle+|1001\rangle)\otimes|1_{\tilde{F}_{2\Omega}}\rangle}{2(1-e^{-2\pi\Omega/a})^{1/2}}\\
&+e^{-\pi\Omega/a} \nu^2\frac{(|0001\rangle+|0010\rangle+|0100\rangle+|1000\rangle)\otimes|1_{\tilde{F}_{1\Omega}}1_{\tilde{F}_{2\Omega}}\rangle}{4(1-e^{-2\pi\Omega/a})} + \mathcal{O}\left( \nu^3 \right).
\end{split}
\end{eqnarray}
In order to determine the dynamical state of the detectors after their interaction with the field, we perform a trace over the external field degrees of freedom, which gives the reduced density matrix for the four-qubit state
\begin{eqnarray}\label{S5}
\begin{split}
\rho^{ABCD(2)}_{\infty(W)}&={\rm{tr}}_{\phi}|W^{(2)}_{4\infty}\rangle\langle W^{(2)}_{4\infty}| + \mathcal{O}\left( \nu^4 \right)\\
&=\rho_{diag}^{W_{4}^{(2)}}+\rho_{nondiag}^{W_{4}^{(2)}}+ \mathcal{O}\left( \nu^{4} \right),
\end{split}
\end{eqnarray}
with
\begin{eqnarray}\label{S7}
\begin{split}
\rho_{diag}^{W_{4}^{(2)}}=&\tilde{K_{0}}|0000\rangle\langle0000|+\tilde{K_{1}}|0001\rangle\langle0001|+\tilde{K_{3}}(|0010\rangle\langle0010|+|0100\rangle\langle0100|\\
&+|1000\rangle\langle1000|)+\tilde{K_{4}}(|0011\rangle\langle0011|+|0101\rangle\langle0101|+|1001\rangle\langle1001|),
\end{split}
\end{eqnarray}
and
\begin{eqnarray}\label{S8}
\begin{split}
\rho_{nondiag}^{W_{4}^{(2)}}=&\tilde{K_{2}}(|0001\rangle\langle0010|+|0001\rangle\langle0100|+|0001\rangle\langle1000|)+\tilde{K_{3}}(|0010\rangle\langle0100|\\
&+|0010\rangle\langle1000|+|0100\rangle\langle1000|)+\tilde{K_{4}}(|0011\rangle\langle0101|+|0011\rangle\langle1001|\\
&+|0101\rangle\langle1001|)+(H.c.)_{nondiag.},
\end{split}
\end{eqnarray}
where $\tilde{K_{0}}=\frac{\nu^{2}}{4(1-q)}$,
$\tilde{K_{1}}=\frac{1-q-\nu^{2}}{4(1-q)}$,
$\tilde{K_{2}}=\frac{2(1-q)-(1+q)\nu^{2}}{8(1-q)}$,
$\tilde{K_{3}}=\frac{1-q-q\nu^{2}}{4(1-q)}$,
and $\tilde{K_{4}}=\frac{q\nu^{2}}{4(1-q)}$.
In contrast to the first-order perturbative treatment, which requires explicit normalization of the reduced density matrix, the second-order framework ensures that the sum of the diagonal elements automatically satisfies the normalization condition $\mathrm{Tr}\rho^{ABCD(2)}_{\infty(W)}=1$, thereby eliminating the necessity for additional normalization procedures.

\begin{figure}
\begin{minipage}[t]{0.5\linewidth}
\centering
\includegraphics[width=3.0in,height=5.2cm]{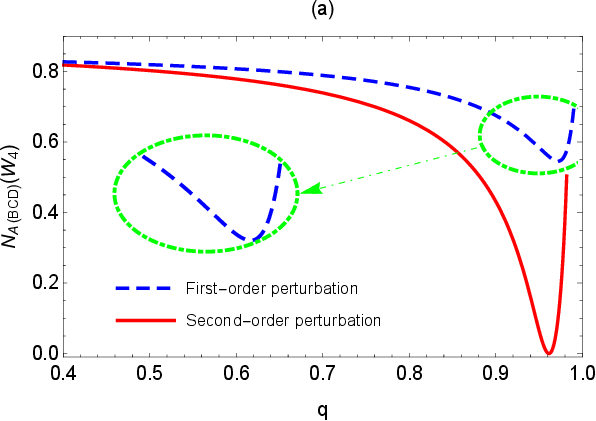}
\label{fig4a}
\end{minipage}%
\begin{minipage}[t]{0.5\linewidth}
\centering
\includegraphics[width=3.0in,height=5.2cm]{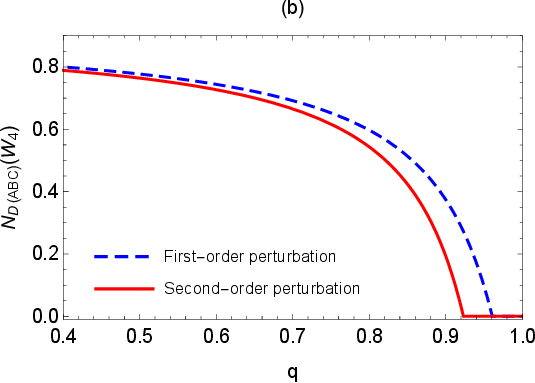}
\label{fig4b}
\end{minipage}%

\begin{minipage}[t]{0.5\linewidth}
\centering
\includegraphics[width=3.0in,height=5.2cm]{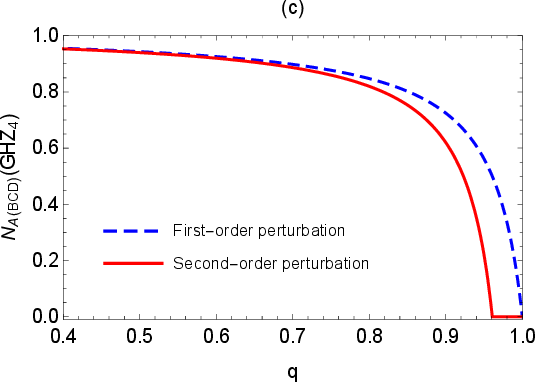}
\label{fig4c}
\end{minipage}%
\begin{minipage}[t]{0.5\linewidth}
\centering
\includegraphics[width=3.0in,height=5.2cm]{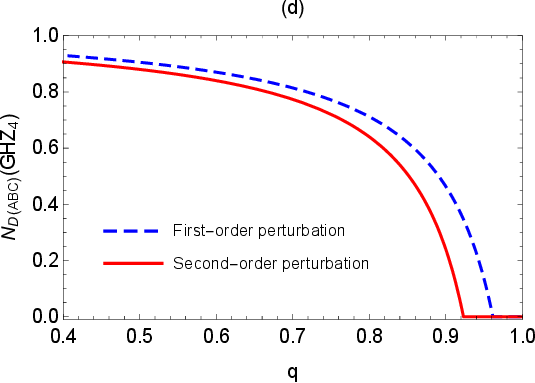}
\label{fig4d}
\end{minipage}%
\caption{The $N_{A(BCD)}$ and $N_{D(ABC)}$ of $W$ and $GHZ$ states as a function of the acceleration parameter $q$ under the first-order perturbation and the second-order perturbation. The effective coupling parameter $\nu$ is fixed as $\nu^{2}=0.04$. }
\label{Fig4}
\end{figure}

In the following, we present the derivation of the $GHZ$ state in the evolution of the tetrapartite Unruh-DeWitt detector system under the second-order perturbation theory.
In the interaction picture, considering the second-order perturbation as given in Eq.(\ref{S1}), the final state of the total system, expressed in terms of the Rindler operators $a^{\rm \dagger}_{RI}$ and $a_{RI}$, can be written as
\begin{eqnarray}\label{SS1}
\begin{split}
|{GHZ}^{(2)}_{4\infty}\rangle=&[I+a^{\rm \dagger}_{RI}(\lambda)D-a_{RI}(\bar{\lambda})D^{\rm \dagger}-\frac{1}{2}(a^{\rm \dagger}_{RI}(\lambda)a_{RI}(\bar{\lambda})DD^{\rm \dagger}\\
&+a_{RI}(\bar{\lambda})a^{\rm \dagger}_{RI}(\lambda)D^{\rm \dagger}D)]|{GHZ}_{4-\infty}\rangle + \mathcal{O}\left( \nu^3 \right).
\end{split}
\end{eqnarray}
By substituting the initial state $|{GHZ}_{4-\infty}\rangle$ from Eq.(\ref{a7}) into Eq.(\ref{SS1}), we obtain
\begin{eqnarray}\label{SS2}
\begin{split}
|{GHZ}^{(2)}_{4\infty}\rangle=&|{GHZ}_{4-\infty}\rangle+\frac{1}{\sqrt{2}}[|1110\rangle\otimes(a^{\rm \dagger}_{RI}(\lambda)|0_{M}\rangle)+|0001\rangle\otimes(a_{RI}(\bar{\lambda})|0_{M}\rangle)+\\
&\frac{1}{2}|0000\rangle\otimes(a^{\rm\dagger}_{RI}(\lambda)a_{RI}(\bar{\lambda})|0_{M}\rangle)+\frac{1}{2}|1111\rangle\otimes(a_{RI}(\bar{\lambda})a^{\rm\dagger}_{RI}(\lambda)|0_{M}\rangle)] + \mathcal{O}\left( \nu^3 \right).
\end{split}
\end{eqnarray}
Using the Bogoliubov transformations provided in Eqs.(\ref{qq3}) and (\ref{qq5}), along with the relations: $a_{M}|0_{M}\rangle=0$ and $a_{M}^{\rm \dagger}|0_{M}\rangle=|1_{M}\rangle$, Eq.(\ref{SS2}) can be rewritten as
\begin{eqnarray}\label{SS3}
\begin{split}
|{GHZ}^{(2)}_{4\infty}\rangle=&|{GHZ}_{4-\infty}\rangle+\frac{1}{\sqrt{2}}\nu[\frac{|1110\rangle\otimes|1_{\tilde{F}_{1\Omega}}\rangle}{(1-e^{-2\pi\Omega/a})^{1/2}}
+e^{-\pi\Omega/a}\frac{|0001\rangle\otimes|1_{\tilde{F}_{2\Omega}}\rangle}{(1-e^{-2\pi\Omega/a})^{1/2}}\\
&+\nu\frac{|0000\rangle\otimes(e^{-\pi\Omega/a}|1_{\tilde{F}_{1\Omega}}1_{\tilde{F}_{2\Omega}}\rangle-e^{-2\pi\Omega/a}|0_{M}\rangle)}{2(1-e^{-2\pi\Omega/a})}\\
&+\nu\frac{|1111\rangle\otimes(e^{-\pi\Omega/a}|1_{\tilde{F}_{1\Omega}}1_{\tilde{F}_{2\Omega}}\rangle-|0_{M}\rangle)}{2(1-e^{-2\pi\Omega/a})}]+ \mathcal{O}\left( \nu^{3} \right)\\
=&\frac{1}{2\sqrt{2}}\bigg[\left(2-\frac{e^{-2\pi\Omega/a} \,\nu^2}{1-e^{-2\pi\Omega/a}}\right)|0000\rangle+\left(2-\frac{\nu^2}{1-e^{-2\pi\Omega/a}}\right)|1111\rangle\bigg]\otimes|0_M\rangle\\
&+\nu\frac{|1110\rangle\otimes|1_{\tilde{F}_{1\Omega}}\rangle}{\sqrt{2}(1-e^{-2\pi\Omega/a})^{1/2}}+e^{-\pi\Omega/a}\nu\frac{|0001\rangle\otimes|1_{\tilde{F}_{2\Omega}}\rangle}{\sqrt{2}(1-e^{-2\pi\Omega/a})^{1/2}}\\
&+e^{-\pi\Omega/a} \nu^2\frac{(|0000\rangle+|1111\rangle)\otimes|1_{\tilde{F}_{1\Omega}}1_{\tilde{F}_{2\Omega}}\rangle}{2\sqrt{2}(1-e^{-2\pi\Omega/a})}+ \mathcal{O}\left( \nu^{3} \right).
\end{split}
\end{eqnarray}
To derive the evolution of the detectors' state after their interaction with the field, we trace out the scalar field degrees of freedom, yielding the reduced density matrix for the four-qubit state
\begin{eqnarray}\label{SS4}
\begin{split}
\rho^{{ABCD}^{(2)}}_{\infty(G)}=&{\rm{tr}}_{\phi}|{GHZ}^{(2)}_{4\infty}\rangle\langle {GHZ}^{(2)}_{4\infty}|+ \mathcal{O}\left( \nu^{4} \right)\\
=&\tilde{L_{0}}|0000\rangle\langle0000|+\tilde{L_{1}}(|0000\rangle\langle1111|+|1111\rangle\langle0000|)+\tilde{L_{2}}|0001\rangle\langle0001|\\
&+\tilde{L_{3}}|1110\rangle\langle1110|+\tilde{L_{4}}|1111\rangle\langle1111|+ \mathcal{O}\left( \nu^4 \right),
\end{split}
\end{eqnarray}
where
$\tilde{L_{0}}=\frac{1-q-q\nu^{2}}{2(1-q)}$,
$\tilde{L_{1}}=\frac{2(1-q)-(1+q)\nu^{2}}{4(1-q)}$,
$\tilde{L_{2}}=\frac{q\nu^{2}}{2(1-q)}$,
$\tilde{L_{3}}=\frac{\nu^{2}}{2(1-q)}$, and
$\tilde{L_{4}}=\frac{1-q-\nu^{2}}{2(1-q)}$.

Using a similar approach, we can calculate the negativity of the tetrapartite $W$ and $GHZ$ states under the second-order perturbation. As shown in Fig.\ref{Fig4}, we present the behavior of the negativity ($1-3$ tangle) of the tetrapartite $W$ and $GHZ$ states as a function of the acceleration parameter $q$, considering both first-order and second-order perturbations. The effective coupling parameter is fixed at $\nu^{2}=0.04$. Our results indicate that the negativity of the $W$ state is smaller under the second-order perturbation relative to the first-order perturbation. In addition, the negativity of the $GHZ$ state undergoes sudden death at a faster rate under the second-order perturbation compared to the first-order perturbation.
These findings reveal a common trend: the negativity of both the tetrapartite $W$ and $GHZ$ states is lower under the second-order perturbation than under the first-order perturbation.


\begin{thebibliography}{99}
\bibitem{M112}
X. M. Hu, Y. Guo, B. H. Liu, C. F. Li and G. C. Guo, Progress in quantum teleportation, Nature Reviews Physics {\bf5}, 339353 (2023).

\bibitem{M113}
X. M. Hu, C. Zhang, B. H. Liu, Y. Cai, X. J. Ye, Y. Guo, W. B. Xing, C. X. Huang, Y. F. Huang, C. F. Li and G. C. Guo, Experimental high-dimensional quantum teleportation, Phys. Rev. Lett {\bf125}, 230501 (2020).

\bibitem{M114}
A. S. Cacciapuoti, M. Caleffi, R. Van Meter and L. Hanzo, When entanglement meets classical communications: Quantum teleportation for the quantum internet, IEEE Trans. Commun. {\bf68}, 3808 (2020).

\bibitem{M2}
F. L. Yan and X. Q. Zhang, A scheme for secure direct communication using EPR pairs and teleportation, Eur. Phys. J. B {\bf41}, 075 (2004).

\bibitem{M3}
T. Gao, F. L. Yan and Y. C. Li, Optimal controlled teleportation, Europhys. Lett. {\bf84}, 50001 (2008).

\bibitem{M115}
M. Gupta and M. J. Nene, Quantum computing: An entanglement measurement, 2020 IEEE International Conference on Advent Trends in Multidisciplinary Research and Innovation (ICATMRI) {\bf30}, 001006 (2020).

\bibitem{M5}
M. A. Nielsen and I. L. Chuang, Quantum Computation and Quantum Information, Cambridge University Press,  (2000).

\bibitem{M6}
A. Datta, S. T. Flammia and C. M. Caves, Entanglement and the power of one qubit, Phys. Rev. A {\bf72}, 042316 (2005).

\bibitem{M116}
J. Yin, $et$ $al.$, Entanglement-based secure quantum cryptography over 1,120 kilometres, Nature {\bf582}, 501 (2020).

\bibitem{M10}
L. Masanes, Universally composable privacy amplification from causality constraints, Phys. Rev. Lett. {\bf102}, 140501 (2009).

\bibitem{M117}
Z. X. Ji, P. R. Fan, H. G. Zhang and H. Z. Wang, Several two-party protocols for quantum private comparison using entanglement and dense coding, Opt. Commun. {\bf459}, 124911 (2020).

\bibitem{M118}
N. Meher, Scheme for realizing quantum dense coding via entanglement swapping, J. Phys. B: At. Mol. Opt. Phys. {\bf53}, 065502 (2020).

\bibitem{M119}
E. Chitambar and F. Leditzky, On the duality of teleportation and dense coding, IEEE Trans. Inf. Theory {\bf70}, 3529 (2023).

\bibitem{M120}
A. Piveteau, J. Pauwels, E. H{\aa}kansson, S. Muhammad, M. Bourennane and A. Tavakoli, Entanglement-assisted quantum communication with simple measurements, Nat. Commun. {\bf13}, 7878 (2022).

\bibitem{M121}
N. Zou, Quantum entanglement and its application in quantum communication, J. Phys.: Conf. Ser. {\bf1827},  012120 (2021).

\bibitem{M12}
J. M. Renes and M. Grassl, Generalized decoding, effective channels, and simplified security proofs in quantum
key distribution, Phys. Rev. A {\bf74}, 022317 (2006).

\bibitem{M13}
Y. H. Zhou, Z. W. Yu and X. B. Wang, Making the decoy-state measurement-device-independent quantum key distribution practically useful, Phys. Rev. A {\bf93}, 042324 (2016).

\bibitem{M123}
M. Ma, Y. Li and J. Shang, Multipartite entanglement measures: a review, Fundamental Research 19 (2024).

\bibitem{M21}
R. Horodecki, P. Horodecki, M. Horodecki and K. Horodecki, Quantum entanglement, Rev. Mod. Phys. {\bf81}, 865 (2009).

\bibitem{M22}
O. G\"{u}hne and G. T\'{o}th, Entanglement detection, Phys. Rep. {\bf474}, 1 (2009).

\bibitem{M23}
N. Friis, G. Vitagliano, M. Malik and M. Huber, Entanglement certification from theory to experiment, Nat.
Rev. Phys. {\bf1}, 72 (2019).

\bibitem{R1}
I. Fuentes-Schuller and R. B. Mann, Alice Falls into a Black Hole: Entanglement in Non-inertial Frames, Phys. Rev. Lett. {\bf95}, 120404 (2005).

\bibitem{R2}
P. M. Alsing, I. Fuentes-Schuller, R. B. Mann and T. E. Tessier, Entanglement of Dirac fields in noninertial frames, Phys. Rev. A {\bf74}, 032326 (2006).

\bibitem{R3}
D. Ahn, Unruh effect as a noisy quantum channel, Phys. Rev. A {\bf98}, 022308 (2018).

\bibitem{R4}
G. Adesso, I. Fuentes-Schuller and M. Ericsson, Continuous-variable entanglement sharing in non-inertial frames, Phys. Rev. A {\bf76}, 062112 (2007).

\bibitem{R5}
T. Gonzalez-Raya, S. Pirandola and M. Sanz, Satellite-based entanglement distribution and quantum teleportation with continuous variables, Commun. Phys. {\bf7},  126 (2024).

\bibitem{R6}
Q. Pan and J. Jing, Hawking radiation, entanglement, and teleportation in the background of an asymptotically flat static black hole, Phys. Rev. D {\bf78}, 065015 (2008).

\bibitem{R7}
M. M. Du, H. W. Li, S. T. Shen, X. J. Yan, X. Y. Li, L. Zhou, W. Zhong and Y. B. Sheng, Maximal steered coherence in the background of Schwarzschild space-time, Eur. Phys. J. C {\bf84}, 450 (2024).

\bibitem{R8}
W. Liu, C. Wen, J. Wang, Lorentz violation alleviates gravitationally induced
entanglement degradation, J. High Energy Phys. {\bf01}, 184 (2025).

\bibitem{R9}
S. Banerjee, A. K. Alok, S. Omkar and R. Srikanth, Characterization of Unruh channel in the context of open quantum systems, J. High Energy Phys. {\bf02}, 82 (2017).

\bibitem{R10}
S. M. Wu, X. W. Fan, R. D. Wang, H. Y. Wu, X. L. Huang and H. S. Zeng, Does Hawking effect always degrade fidelity of quantum teleportation in Schwarzschild spacetime?, J. High Energy Phys. {\bf11}, 232 (2023).

\bibitem{R11}
J. Le\'{o}n and E. Mart\'{\i}n-Mart\'{\i}nez, Spin and occupation number entanglement of Dirac fields for noninertial observers, Phys. Rev. A {\bf80}, 012314 (2009).

\bibitem{R12}
J. He, S. Xu, Y. Yu and L. Ye, Property of various correlation measures of open Dirac system with Hawking effect in Schwarzschild space-time, Phys. Lett. B {\bf740}, 322 (2015).

\bibitem{R13}
S. Elghaayda, A. Ali, S. Al-Kuwari and M. Mansour, Physically accessible and inaccessible quantum correlations of Dirac fields in Schwarzschild spacetime, Phys. Lett. A {\bf525}, 129915 (2024).

\bibitem{R14}
J. K. Basak, D. Giataganas, S. Mondal and W. Y. Wen, Reflected entropy and Markov gap in noninertial frames, Phys. Rev. D {\bf108}, 125009 (2023).

\bibitem{R15}
S. Sen, A. Mukherjee and S. Gangopadhyay, Entanglement degradation as a tool to detect signatures of modified gravity, Phys. Rev. D {\bf109}, 046012 (2024).

\bibitem{R16}
A. Ali, S. Al-Kuwari, M. Ghominejad, M. T. Rahim, D. Wang and S. Haddadi, Quantum characteristics near event horizons, Phys. Rev. D {\bf110}, 064001 (2024).

\bibitem{R17}
H. Wu and L. Chen, Orbital angular momentum entanglement in noninertial reference frame, Phys. Rev. D {\bf107}, 065006 (2023).

\bibitem{R18}
C. Y. Liu, Z. W. Long and Q. L. He, Quantum coherence and quantum Fisher information of Dirac particles in curved spacetime under decoherence, Phys. Lett. B {\bf857}, 138991 (2024).

\bibitem{RM1}
M. R. Hwang, D. Park and E. Jung, Tripartite entanglement in a noninertial frame, Phys. Rev. A \textbf{83}, 012111 (2011).

\bibitem{RM2}
S. Xu, X. K. Song, J. D. Shi and L. Ye, How the Hawking effect affects multipartite entanglement of Dirac particles in the background of a Schwarzschild black hole, Phys. Rev. D {\bf89}, 065022 (2014).

\bibitem{RM3}
W. C. Qiang, G. H. Sun, Q. Dong and S. H. Dong, Genuine multipartite concurrence for entanglement of Dirac fields in noninertial frames, Phys. Rev. A {\bf98}, 022320 (2018).

\bibitem{RM4}
H. M. Reji, H. S. Hegde and R. Prabhu, Conditions for separability in multiqubit systems with an accelerating qubit using a conditional entropy, Phys. Rev. A {\bf110}, 032403 (2024).

\bibitem{RM5}
A. J. Torres-Arenas, Q. Dong, G. H. Sun, W. C. Qiang and S. H. Dong, Entanglement measures of W-state in noninertial frames, Phys. Lett. B {\bf789}, 93 (2019).

\bibitem{RM6}
S. M. Wu, X. W. Teng, J. X. Li, S. H. Li, T. H. Liu and J. C. Wang, Genuinely accessible and inaccessible entanglement in Schwarzschild black hole, Phys. Lett. B {\bf848}, 138334 (2024).

\bibitem{RM7}
T. Zhang, X. Wang and S. M. Fei, Hawking effect can generate physically inaccessible genuine tripartite nonlocality, Eur. Phys. J. C {\bf83}, 607 (2023).

\bibitem{RM8}
S. M. Wu and H. S. Zeng, Genuine tripartite nonlocality and entanglement in curved spacetime, Eur. Phys. J. C {\bf82}, 4 (2022).

\bibitem{RM9}
L. J. Li, F. Ming, X. K. Song, L. Ye and D. Wang, Quantumness and entropic uncertainty in curved space-time, Eur. Phys. J. C  {\bf82}, 726 (2022).

\bibitem{RM10}
T. Y. Wang and D. Wang, Entropic uncertainty relations in Schwarzschild space-time, Phys. Lett. B {\bf855}, 138876 (2024).

\bibitem{RM11}
S. Harikrishnan, S. Jambulingam, P. P. Rohde and C. Radhakrishnan, Accessible and inaccessible quantum coherence in relativistic quantum systems, Phys. Rev. A {\bf105}, 052403 (2022).

\bibitem{RM12}
W. M. Li and S. M. Wu, Bosonic and fermionic coherence of N-partite states in
the background of a dilaton black hole, J. High Energy Phys. {\bf09}, 144 (2024).

\bibitem{QQRM12}
W. Liu, D. Wu, J. Wang, Light rings and shadows of static black holes in effective quantum gravity, Phys. Lett. B {\bf858}, 139052  (2024).

\bibitem{WDD1}
D. E. Bruschi, J. Louko, E. Mart\'{\i}n-Mart\'{\i}nez, A. Dragan, I. Fuentes,  Unruh effect in quantum information beyond the single-mode approximation, Phys. Rev. A {\bf82}, 042332 (2010).

\bibitem{WDD2}
D. E. Bruschi, A. Dragan, I. Fuentes, J. Louko, Particle and anti-particle bosonic entanglement in non-inertial frames,
Phys. Rev. D {\bf86}, 025026 (2012).

\bibitem{rbn1}
G. W. Mi, X. Huang, S. M. Fei, T. Zhang, Genuine four-partite Bell nonlocality in the curved spacetime, Eur. Phys. J. C  {\bf85}, 354 (2025).










\bibitem{rbn2}
L. J. Li, X. K. Song, L. Ye, and D. Wang, Quantifying quantumness in (A)dS spacetimes with Unruh-DeWitt detector, Phys. Rev. D {\bf111}, 065007 (2025).

\bibitem{SI1}
J. Steinhauer, et al., Analogue cosmological particle creation in an ultracold quantum fluid of light, Nat. Commun. {\bf13}, 2890 (2022).

\bibitem{SI2}
Z. Tian, L. Wu, L. Zhang, J. Jing and J. Du, Probing Lorentz-invariance-violation-induced nonthermal Unruh effect in quasi-two-dimensional dipolar condensates, Phys. Rev. D {\bf106}, L061701 (2022).

\bibitem{SI3}
J. Hu, L. Feng, Z. Zhang and C. Chin, Quantum simulation of Unruh radiation, Nat. Phys. {\bf15}, 785 (2019).

\bibitem{SI4}
Y. H. Shi, et al., Quantum simulation of Hawking radiation
and curved spacetime with a superconducting on-chip black hole, Nat. Commun. {\bf14}, 3263 (2023).

\bibitem{SI5}
J. Drori, Y. Rosenberg, D. Bermudez, Y. Silberberg and U. Leonhardt, Observation of Stimulated Hawking Radiation in an Optical Analogue, Phys. Rev. Lett. {\bf122}, 010404 (2019).

\bibitem{SI6}
I. Agullo, A. J. Brady and D. Kranas, Quantum Aspects of Stimulated Hawking Radiation in an Optical Analog White-Black Hole Pair, Phys. Rev. Lett. {\bf128}, 091301 (2022).

\bibitem{SI7}
Z. Tian, J. Jing and A. Dragan, Analog cosmological particle generation in a superconducting circuit,
Phys. Rev. D {\bf95}, 125003 (2017).

\bibitem{EI1}
H. N. Wu, Y. H. Li, B. Li, X. You, R. Z. Liu, J. G. Ren, J. Yin, C. Y. Lu, Y. Cao, C. Z. Peng and J. W. Pan, Single-Photon Interference over 8.4 km Urban Atmosphere: Toward Testing Quantum Effects in Curved Spacetime with Photons, Phys. Rev. Lett. {\bf133}, 020201 (2024).

\bibitem{EI2}
P. Xu et al., Satellite testing of a gravitationally induced quantum decoherence model, Science {\bf366}, 132 (2019).

\bibitem{EI3}
C. W. Chou, D. B. Hume, T. Rosenband and D. J. Wineland, Optical Clocks and Relativity, Science {\bf329}, 1630 (2010).

\bibitem{EI4}
J. Y. Wang et al., Direct and full-scale experimental verifications towards ground-satellite quantum key distribution, Nat. Photonics {\bf7}, 387 (2013).

\bibitem{UD1}
W. G. Unruh, Notes on black-hole evaporation, Phys. Rev. D {\bf14}, 870 (1976).

\bibitem{UD2}
B. S. DeWitt, Quantum Gravity: The New Synthesis, Cambridge University Press, (1979).

\bibitem{UD3}
N. D. Birrell, P. C. W. Davies, Quantum Fields in Curved Space, Cambridge Monographs on Mathematical
Physics, Cambridge University Press, (1984).

\bibitem{UD4}
S. W. Hawking and W. Israel, General Relativity: An Einstein Centenary Survey, Cambridge University Press (1980).

\bibitem{UD5}
B. Reznik, A. Retzker and J. Silman, Violating Bell's inequalities in vacuum, Phys. Rev. A {\bf71}, 042104 (2005).

\bibitem{UD6}
L. C. C\'{e}leri, A. G. S. Landulfo, R. M. Serra and G. E. A. Matsas, Sudden change in quantum and classical
correlations and the Unruh effect, Phys. Rev. A {\bf81}, 062130 (2010).

\bibitem{UD7}
D. M. Avalos, K. Gallock-Yoshimura, L. J. Henderson and R. B. Mann, Instant extraction of nonperturbative tripartite entanglement, Phys. Rev. Research {\bf5}, L042039  (2023).

\bibitem{UD8}
T. R. Perche, J. Polo-G\'{o}mez,  B. de S. L. Torres and E. Mart\'{\i}n-Mart\'{\i}nez, Fully relativistic entanglement harvesting, Phys. Rev. D {\bf109}, 045018 (2024).

\bibitem{UD9}
Z. Liu, J. Zhang and H. Yu, Harvesting correlations from vacuum quantum felds in the presence of a refecting boundary, J. High Energy Phys. {\bf11}, 184 (2023).

\bibitem{UD10}
Y. Ji, J. Zhang and H. Yu, Entanglement harvesting in cosmic string spacetime, J. High Energy Phys. {\bf06}, 161 (2024).

\bibitem{UD11}
Q. Liu, T. Liu, C. Wen, J. Wang, Optimal quantum strategy for locating Unruh channels, Phys. Rev. A {\bf110}, 022428 (2024).

\bibitem{UD12}
K. Gallock-Yoshimura and R. B. Mann, Entangled detectors nonperturbatively harvest mutual information, Phys. Rev. D {\bf104}, 125017 (2021).

\bibitem{UD13}
S. M. Wu, R. D. Wang, X. L. Huang and Z. Wang, Does gravitational wave assist vacuum steering and Bell nonlocality?, J. High Energy Phys. {\bf07}, 155 (2024).

\bibitem{UD14}
A. G. S. Landulfo and G. E. A. Matsas, Sudden death of entanglement and teleportation fidelity loss via the Unruh effect, Phys. Rev. A {\bf80}, 032315 (2009).

\bibitem{UD15}
M. M. Du, H. W. Li, Z. Tao, S. T. Shen, X. J. Yan, X. Y. Li, W. Zhong, Y. B. Sheng and L. Zhou, Basis-independent quantum coherence and its distribution under relativistic motion, Eur. Phys. J. C {\bf84}, 838 (2024).

\bibitem{UD16}
J. Wang, Z. Tian, J. Jing and H. Fan, Irreversible degradation of quantum coherence under relativistic motion, Phys. Rev. A {\bf93}, 062105 (2016).

\bibitem{UD17}
Z. Tian, J. Wang and J. Jing, Nonlocality and entanglement via the Unruh effect, Ann. Phys. {\bf332}, 98 (2012).

\bibitem{UD18}
T. Liu, J. Wang, J. Jing and H. Fan, The influence of Unruh effect on quantum steering for accelerated two-level detectors with different measurements, Ann. Phys. {\bf390}, 334 (2018).

\bibitem{UD19}
Y. K. Zhang, L. J. Li, X. K. Song, L. Ye and D. Wang, Entropic uncertainty and quantum non-classicality of Unruh-Dewitt detectors in relativity, Phys. Lett. B {\bf858}, 139063 (2024).

\bibitem{UD20}
J. He, Z. Y. Ding, J. D. Shi and T. Wu, Multipartite quantum coherence and distribution under the Unruh effect, Ann. Phys. {\bf530}, 1800167 (2018).

\bibitem{NE1}
Y. C. Ou and H. Fan, Monogamy inequality in terms of negativity for three-qubit states, Phys. Rev. A {\bf75}, 062308 (2007).



\bibitem{NE2}
R. H. Jonsson, E. Mart\'{\i}n-Mart\'{\i}nez, and A. Kempf, Quantum signaling in cavity QED, Phys. Rev. A {\bf89}, 022330 (2014).

\bibitem{NE3}
R. H. Jonsson, Quantum signaling in relativistic motion and across acceleration horizons, J. Phys. A: Math. Theor. {\bf50}, 355401 (2017).










\end{thebibliography}
\end{document}